\documentclass[useAMS,usenatbib]{mn2e}
\usepackage{float,epsfig}
\usepackage{pstricks}

\newcommand{\mincir}{\raise
  -2.truept\hbox{\rlap{\hbox{$\sim$}}\raise5.truept \hbox{$<$}\ }}
\newcommand{\magcir}{\raise
  -2.truept\hbox{\rlap{\hbox{$\sim$}}\raise5.truept \hbox{$>$}\ }}

\title[XMM-Newton study of groups. I.]{XMM-Newton study of $0.012<z<0.024$
  groups. I: Overview of the IGM thermodynamics.}

\author[Finoguenov et al.]{A. Finoguenov$^{1,2}$\thanks{E-mail:
alexis@mpe.mpg.de}, T.J. Ponman$^3$, J.P.F. Osmond$^{3,4}$,
  M. Zimer$^{1,5}$\\
{$^1$ Max-Planck-Institut f\"ur extraterrestrische Physik,
             Giessenbachstra\ss e, 85748 Garching, Germany}\\
{$^2$ Space Research Institute, Profsoyuznaya 84/32, Moscow, 117810, Russia}\\
{$^3$ School of Physics \& Astronomy, University of Birmingham,
  Edgbaston, Birmingham B15 2TT, UK}\\
{$^4$ Centre for Electronic Imaging, School of Engineering and Design,
Brunel University, Uxbridge, UB8 3PH, UK}\\
{$^5$ Institut fuer Astronomie, Universitaet Wien,
Tuerkenschanzstrasse 17, 1180 Wien, Austria}}
\begin{document}
\date{in preparation for MNRAS, last revised \today}

\pagerange{\pageref{firstpage}--\pageref{lastpage}} \pubyear{2006}

\maketitle

\label{firstpage}

\begin{abstract}
We study the thermodynamic properties of the hot gas in a sample of groups
in the 0.012--0.024 redshift range, using XMM-Newton observations. We
present measurements of temperature, entropy, pressure and
iron abundance. Non-parametric fits are used to derive the mean
properties of the sample and to study dispersion in the values of entropy 
and pressure. The scaling of the entropy at $0.2r_{500}$ matches well the
results of Ponman et al. (2003). However, compared to cool clusters, the
groups in our sample reveal larger entropy at inner radii and a
substantially flatter slope in the entropy in the outskirts, compared to
both the prediction of pure gravitational heating and to observations of
clusters. This difference corresponds to the systematically flatter 
group surface brightness profiles, reported previously. The scaled 
pressure profiles can be well approximated with a S\'ersic model 
with $n=4$. We find that groups exhibit a systematically larger 
dispersion in pressure, compared to clusters of galaxies, while the 
dispersion in entropy is similar.
\end{abstract}


\section{Introduction}

Use of clusters of galaxies in cosmological studies, requires a relation
between the observed properties of the systems, such as X-ray luminosity,
ICM temperature and the total mass of the system. Understanding this
relation and its evolution with the redshift is therefore one of the key
research focuses of modern astrophysics (e.g. Majumdar \& Mohr 2004).  Early
observations have already demonstrated the importance of processes related
to the physics of baryons in defining the observational appearance of
clusters, and in particular on their low-mass end -- groups of galaxies
(e.g. Ponman et al. 1999). Currently, theoretical interpretation of the
observed properties of groups and clusters of galaxies involves an interplay
between the cooling and resulting star-formation and feedback (Borgani et
al. 2001; Finoguenov et al. 2003; Voit et al. 2003; Kay et al. 2004).  The
complexity of the feedback schemes and variations in the feedback
efficiency, given by IMF as well as AGN activity (e.g. Springel \& Hernquist
2003), points to the importance of observations. In particular, observations
of the cores of groups and clusters of galaxies are critical to understand
the effects of cooling (Voit et al. 2001).  Finoguenov et al. (2002)
demonstrated that outskirts of groups and cool clusters also deviate from
the expectations of gravitational heating, and cannot be explained by the
effects of cooling. The challenging energetics of the observed effect,
requires a very efficient feedback scheme, and the currently favoured
mechanism invokes amplification of the entropy of the accreting gas by the
shock heating -- so called ``smooth accretion'' (Ponman et al. 2003; Voit et
al. 2003; Voit \& Ponman 2003; Borgani et al. 2005).

The large grasp of XMM-Newton provides us with new possibilities to test
scenarios of group formation -- in particular, the link between the state of
the gas and the structure in X-ray images. Our idea here is quite
straight-forward; since mergers are examples of lumpy accretion, post-merger
groups should have on average lower entropy, compared to systems where the
bulk of the material has been added through slow accretion. Thus, using two
dimensional analysis of XMM-Newton data, combined with standard analysis
using annuli, with limiting radii increasing logarithmically, we can both
infer the state of the gas with unprecedented precision and provide an
explanation of the observed trend by studying substructure. We also examine
the dynamic state of the galaxies in these groups, to provide independent
evidence.

At the moment there is no large purely X-ray selected sample of local groups
of galaxies. For example, only a few objects, Fornax cluster, MKW4, NGC4636,
NGC1550 and NGC5044 are present in the HIFLUGS, a complete all-sky sample of
brightest groups and clusters of galaxies (Reiprich \& B\"ohringer 2002).
Most present-day samples of groups are based on the X-ray follow-up of the
optical surveys (e.g. Mahdavi et al. 2000; Mulchaey et al. 2003). For this
study we have primarily selected the groups primarily from Mulchaey et
al. (2003) with publicly available XMM-Newton (Jansen et al. 2001)
observations. Mulchaey list is based on the cross-correlating the ROSAT
observation log with the positions of optically-selected groups in the
catalogs of Huchra \& Geller (1982), Geller \& Huchra (1983), Maia et
al. (1989), Nolthenius (1993) and Garcia (1993). While these catalogs also
include richer galaxy systems (i.e. clusters), the Mulchaey list only
includes systems with velocity dispersions less than 600 km/s or an
intragroup medium temperature less than 2 keV.

A total of 25 systems observed by XMM-Newton have been selected, and further
divided onto two samples covering the redshift ranges 0.004--0.012 and
0.012--0.024. The low-redshift sample is discussed in Finoguenov et al.
(2006), while here we concentrate on the high redshift sample.  Although our
sample is not statistically complete, it has been shown by Finoguenov et al.
(2006) that both XMM-Newton follow-up and the selection of the groups by
Mulchaey et al. (2003) appear representative. We further increase our
$z=$0.012--0.024 sample by addition of two groups, HCG51 and Pavo, which
were not in Mulchaey's sample (there has been no ROSAT PSPC observations of
those two groups), but lie in the same redshift range.

An important difference between our `low' and `high' redshift subsamples
consists in the different coverage of the group emission.  While in the
low-redshift sample, we concentrate on the properties of the central
brightest group galaxies (BGGs), with the high-redshift sample typical
scales resolved correspond to the transition zone between the BGG and the
group. As the Mulchaey list is flux-limited, the typical luminosity of the
high-redshift subsample is $10^{42}$ ergs s$^{-1}$, i.e. we look at bona
fide groups, where the total X-ray flux is dominated by the group emission,
while the origin of the hot gas is a specific aspect of the low-z subsample.
In present sample, we have, however, included a few examples (e.g. HCG92),
where strong galaxy interactions result in the high X-ray luminosity.

The present paper (Paper I) presents an analysis of the radial and
2-dimensional structure of the hot gas properties for the sample.  The paper
is organized as follows: \S\ref{s:data} describes the analysis of the
XMM-Newton observations; \S\ref{s:results} outlines the average properties
of the sample and dispersion around the mean trends; \S\ref{s:ind} describes
each group of the sample individually; \S\ref{s:sum} concludes the paper.

Paper II (Osmond et al. 2006) is concerned with classifying the groups, on
the basis of their gas properties, and looking at the relationship between
different properties. Finally, a study of temperature and element abundance
profiles using the Chandra observations of 15 groups with 6 overlapping
with the current sample are reported in Rasmussen et al. (2006). It turns
out that the Chandra sample is dominated by the cool core systems, which
results in some differences in the reported mean trends. However for
systems in common the results agree well.


\section{Data}\label{s:data}

Tab.\ref{t:ol} details the observations, listing the name of the group
(column 1), the assigned XMM archival name (2), net Epic-pn exposure after
removal of flaring episodes (3), pn filter used (4), needed for instrumental
response as well as background estimates, XMM-Newton revolution number (5),
useful assessment to the secular evolution of the instrumental background,
pn frame time (6), which determines the fraction of out-of-time events.

\begin{table}
\begin{center}
\renewcommand{\arraystretch}{1.1}\renewcommand{\tabcolsep}{0.12cm}
\caption{Log of the XMM Epic-pn observations of groups.}
\label{t:ol}

\begin{tabular}{rccccr}
\hline
 Name &  Obs.       &   net    &  pn     & XMM   &frame\\
      &   ID        &   exp.    &  Filter & Orbit &time\\
      &             &   ksec    &         &       &ms\\
\hline
3C449   & 0002970101& 16.1&Medium & 367&  73 \\
HCG 92  & 0021140201& 30.1&Thin   & 366&  73 \\
Pavo    & 0022340101&  9.1&Thin   & 423&  73 \\
HCG 42  & 0041180301& 16.4&Medium & 358&  73 \\
HCG 68  & 0041180401& 16.3&Thick  & 454&  73 \\
NGC 5171& 0041180801& 13.2&Medium & 377& 199 \\
HCG 15  & 0052140301& 22.9&Thin   & 383& 199 \\
NGC 507 & 0080540101& 25.9&Thin   & 202& 199 \\
NGC 4073& 0093060101&  9.4&Medium & 373& 199 \\
NGC 4325& 0108860101& 14.8&Thin   & 191&  73 \\
NGC 2563& 0108860501& 15.9&Medium & 339&  73 \\
NGC 533 & 0109860101& 28.8&Thin   & 195& 199 \\
HCG51   & 0112270301&  5.4&Thin   & 363& 199 \\
HCG 62  & 0112270701&  8.1&Medium & 568& 199 \\
\hline
\end{tabular}
\end{center}
\end{table}

Initial steps of data reduction are similar to the procedure described in
Zhang et al. (2003), Finoguenov et al. (2003). As most of the observations
analyzed here were performed using a short integration frame time for pn, it
is important to remove the out-of-time events from imaging and spectral
analysis. We used the standard product of {\it epchain} SAS task to produce
the simulated OOTE file for all the observations and scale it by the
fraction of the OOTE expected for the frame exposure time, as specified in
Tab.\ref{t:ol}.  For further details of XMM-Newton processing we refer the
reader to {\it http://wave.xray.mpe.mpg.de/xmm/cookbook/general} page.

The analysis consists of two parts: first, revealing the structure in the
surface brightness and temperature maps and second, verifying it through the
spectral analysis. The first part consists in producing temperature
estimates, based on the calibrated wavelet prefiltered hardness ratio maps
and producing the projected pressure and entropy maps. Wavelet filtering
(Vikhlinin et al. 1998) is used to find the structure and control its
significance.

The second, spectroscopy part of the analysis, uses the mask file, created
based on the results of both hardness ratio and surface brightness analysis
described above. First application of this technique is presented in
Finoguenov et al. (2003).

The approach to the analysis of the sample is as follows. Taking the
directly observed image we divide it along the contours of surface
brightness. For a symmetrical system, the result of this procedure will be
annuli. For a system in hydrostatic equilibrium, and neglecting the effect
of Fe abundance variations on the emission, such selection alone is
sufficient to remove possible temperature mixing (the X-ray theorem, Buote
\& Canizares 1994, which although strictly valid in 3D, could be shown to
result in a symmetry in a projection onto the observer's plane for a single
triaxial potential; the case of multiple subhalos will, however, suffer from
projection effects). Next, we study the hardness ratio map, which is
sensitive to temperature variations, allowing us to avoid mixing multiple
temperature components in a blind spectral extraction. In combination with
the first step, we are looking for deviations from hydrostatic equilibrium,
or a presence of subhalos.  We also build the projected entropy and pressure
maps using an image and the hardness ratio map.  Although this way of map
construction suffers from a number of degeneracies, with metalicity-density
being the strongest (a significant fraction of the group emission is due to
the line emission), these maps indicate the regions of primary interest for
detailed spectroscopic analysis, in which most of the degeneracies are
removed. In general, the entropy should monotonically increase with the
increasing radius, while the pressure should decrease.

The spectral analysis was performed using a single-temperature APEC plasma
code with solar abundance pattern. Absorption was fixed at the galactic
value, reported in Tab.\ref{t:basic}. In this work we employ a very detailed
modeling of the background. First we consider the outermost part of the
detector in respect to the group centre, in order to estimate the
instrumental background. We subtract the background accumulation in order to
remove most of the background and later look for the residuals. In our
analysis we used the Read \& Ponman (2003) background accumulation for the
medium filter, XMM observation of CDFS (Streblyanska et al. 2006; Finoguenov
in prep.) for the observations performed with the thin filter and an
accumulation by Rossetti et al. (2005) for the thick filter.  We use
0.45--7.5 and 9.5--12 keV bands and test the presence of the following
components: hard instrumental background, having a slope of around 0. This
background dominates at energies above 5 keV.  Such a component is found in
observations 0041180301 and 0041180801. Another component has a steeper
index, 1--3, and is associated with the soft proton component. Such a
component was added to the background for observations 0041180301,
0041180801 and 0112270301. Our analysis demonstrated that the scaling of
these components, in particular the soft proton component, is somewhat more
complicated than equal flux per detector area, we left the normalizations of
these components free. We point out that an advantage of working with
low-redshift groups, compared to clusters, is that group emission from a
$\sim$1 keV thermal plasma is very different from the power law shape of the
background.

Once the background components are estimated, we switch to the narrow energy
band, 0.45--3 keV for fitting the source. We also consider adding a thermal
component with a fixed temperature of 0.2 keV to account for the possible
variation in the galactic foreground emission. This component was found to
be significant in observations 0021140201 (HCG 92), 0022340101 (Pavo),
0041180801 (NGC 5171), 0108860101 (NGC 4325) and 0112270701 (HCG 51). In
0041180801, the temperature of this component was significantly higher,
$0.35\pm0.04$ keV. In the spectral analysis we group the channels to achieve
30 counts per bin, allowing us to apply $\chi^2$ fitting.  All identified
point sources were excised from the spectral extraction.

In calculating the local gas properties, we perform an estimate of the
projection length of each analyzed region to obtain actual gas properties at
these locations, as described at length in Henry et al. (2004) and Mahdavi
et al. (2005). Since we do not perform a deprojection of the spectra, to
reduce the importance of the projection effects we discard regions having a
ratio of the minimal to the maximal distance from the centre of the group of
values exceeding 0.8. The centre of a group is taken at the peak of the
extended X-ray emission.

\section{Average properties of the sample}\label{s:results}

We start this section by examining the mean trends seen in the data,
comparing to previous results and then looking at the individual properties
of the groups, where we identify the underlying cause for deviation of
individual systems from the mean trend.

Table \ref{t:basic} lists the known properties of the groups. Column (1)
identifies the system, (2) gives the value of galactic absorption towards
the group (in spectral fitting we freeze the value of the absorbing column
to this value), (3) redshift of the group, (4) velocity dispersion of the
group galaxies, calculated as in Osmond \& Ponman (2004), (5) optical
luminosity of the group and separately of brightest group galaxy from Osmond
\& Ponman (2004), (6-7) temperature in the range $0.1-0.3r_{500}$ used for
scaling ($T_w$), and estimate of $r_{500}$, obtained iteratively.  The
$r_{500}$ (radius within which the mean density is 500 times the critical
value) is calculated as $r_{500}=0.391 {\rm Mpc} \times (kT_w/{\rm
keV})^{0.63}h_{70}^{-1}$ using the M--T relation (Pacaud, F. 2005, private
communication) rederived from the Finoguenov et al. (2001) using an
orthogonal regression and correcting the masses to $h_{70}$ and a LCDM
cosmology. The $h(z)=(\Omega_M (1+z)^3 + \Omega_\Lambda)^{1/2}$ correction,
is negligible for this sample and has not been applied. The suggested
modified entropy scaling goes as $S \propto T_w^{2/3}$ (Ponman et al. 2003),
which corresponds to a scaling of pressure $P \propto T_w^{3/2}$.

\begin{table}
\begin{center}
\renewcommand{\arraystretch}{0.9}\renewcommand{\tabcolsep}{0.04cm}
\caption{Basic properties of groups.}
\label{t:basic}
\begin{tabular}{lcccccc}
\hline
\multicolumn{2}{l}{Group \ \ \ \ $N_H$} & z & $\sigma_{group}$&  $L_B$
(BGG)&  kT$_w$ & $r_{500}$ \\
\multicolumn{2}{c}{\hspace*{0.9cm}  $10^{20}$ cm$^{-2}$}&
&km/s&$10^{10}L_\odot$& keV & kpc \\ 
\hline               
3C449   & 11.8&0.0171&$335\pm112$& 7.8 (2.24) & $1.28\pm 0.02$&453\\ 
HCG92   & 8.03&0.0221&$467\pm176$&11.5 (3.31) & $0.79\pm 0.24$&334\\ 
Pavo    & 5.18&0.0134&$440\pm96 $&28.1 (8.84) & $0.77\pm 0.12$&330\\ 
HCG42   & 4.78&0.0132&$282\pm43 $&21.3 (8.91) & $0.75\pm 0.19$&324\\ 
HCG68   & 0.97&0.0078&$191\pm68 $&25.7 (4.37) & $0.69\pm 0.09$&308\\ 
NGC5171 & 1.93&0.0231&$494\pm99 $&19.1 (5.75) & $1.21\pm 0.05$&436\\ 
HCG15   & 3.20&0.0225&$404\pm122$& 7.1 (1.82) & $0.62\pm 0.04$&286\\ 
NGC507  & 5.24&0.0165&$580\pm94 $&48.8 (4.05) & $1.34\pm 0.01$&467\\ 
NGC4073 & 1.89&0.0199&$565\pm72 $&50.1 (15.5) & $1.87\pm 0.05$&575\\ 
NGC4325 & 2.23&0.0257&$376\pm70 $&11.5 (4.07) & $1.01\pm 0.01$&389\\ 
NGC2563 & 4.23&0.0149&$384\pm49 $&28.2 (4.27) & $1.31\pm 0.05$&460\\ 
NGC533  & 3.10&0.0185&$439\pm60 $&33.1 (9.77) & $1.26\pm 0.01$&448\\ 
HCG51   & 1.27&0.0265&$546\pm151$&17.4 (3.59) & $1.16\pm 0.13$&424\\ 
HCG62   & 3.01&0.0145&$418\pm51 $&31.6 (3.47) & $1.06\pm 0.02$&403\\ 
\hline               
\end{tabular}
\end{center}
\end{table}

Tables \ref{t:prop1}--\ref{t:prop3} present the characteristics of the
groups obtained by mass-averaging of the observed spectroscopic components
separated using the following radial bins: $<0.1r_{500}$, $0.1-0.3r_{500}$,
$0.3-0.7r_{500}$. Col. (1) identifies the group, (2) reports the temperature
in keV, (3) iron abundance as a fraction of the photospheric solar value of
Anders \& Grevesse (1989), (4) entropy, (5) pressure.

\begin{table}
\begin{center}
\renewcommand{\arraystretch}{0.9}\renewcommand{\tabcolsep}{0.05cm}
\caption{Properties of groups within $0.1 r_{500}$.}
\label{t:prop1}

\begin{tabular}{rccccccccccc}
\hline
 Name & kT  & Z & S & P \\
      & keV & $Z_\odot$&keV cm$^2$ & $10^{-12}$ dyne cm$^{-2}$ \\
\hline
3C449   &$1.38\pm 0.02$&$0.33\pm 0.03$ & $  61.\pm 1.8$ & $  7.48\pm 0.30$ \\
HCG92   &$0.69\pm 0.05$&$0.22\pm 0.04$ & $  28.\pm 4.1$ & $  6.20\pm 0.77$ \\
Pavo    &$0.98\pm 0.08$&$0.25\pm 0.07$ & $  54.\pm 6.2$ & $  4.23\pm 0.50$ \\
HCG42   &$0.81\pm 0.07$&$0.30\pm 0.03$ & $  49.\pm 6.6$ & $  3.01\pm 0.39$ \\
HCG68   &$0.66\pm 0.02$&$0.18\pm 0.04$ & $  80.\pm10.2$ & $  0.88\pm 0.14$ \\
NGC5171 &$1.55\pm 0.10$&$0.21\pm 0.09$ & $ 111.\pm10.2$ & $  4.46\pm 0.47$ \\
HCG15   &$0.66\pm 0.13$&$0.01\pm 0.02$ & $  34.\pm 8.2$ & $  2.58\pm 0.72$ \\
NGC507  &$1.27\pm 0.01$&$0.58\pm 0.02$ & $  42.\pm 0.6$ & $ 11.06\pm 0.22$ \\
NGC4073 &$1.96\pm 0.04$&$0.82\pm 0.07$ & $  70.\pm 2.1$ & $ 15.59\pm 0.52$ \\
NGC4325 &$0.94\pm 0.01$&$0.53\pm 0.06$ & $  24.\pm 1.3$ & $ 11.49\pm 0.93$ \\
NGC2563 &$1.52\pm 0.04$&$0.44\pm 0.09$ & $  99.\pm 6.0$ & $  4.87\pm 0.39$ \\
NGC533  &$1.32\pm 0.01$&$0.50\pm 0.03$ & $  70.\pm 1.6$ & $  5.68\pm 0.18$ \\
HCG51   &$1.18\pm 0.15$&$0.59\pm 0.06$ & $  92.\pm19.9$ & $  2.99\pm 0.59$ \\
HCG62   &$1.13\pm 0.02$&$0.37\pm 0.04$ & $  46.\pm 2.0$ & $  7.46\pm 0.40$ \\
 \hline
\end{tabular}
\end{center}

\begin{center}
\renewcommand{\arraystretch}{0.9}\renewcommand{\tabcolsep}{0.05cm}
\caption{Properties of groups between $0.1 r_{500}$ and $0.3 r_{500}$.}
\label{t:prop2}

\begin{tabular}{rccccccccccc}
\hline
 Name & kT  & Z & S & P \\
      & keV & $Z_\odot$&keV cm$^2$ & $10^{-12}$ dyne cm$^{-2}$ \\
\hline
3C449 & $1.05\pm 0.02$ & $0.09\pm 0.01$ & $112\pm  3$ & $1.48\pm 0.04$\\
HCG92 & $0.78\pm 0.30$ & $0.50\pm 0.93$ & $251\pm143$ & $0.21\pm 0.16$\\
Pavo  & $0.66\pm 0.08$ & $0.36\pm 0.06$ & $106\pm 19$ & $0.51\pm 0.10$\\
HCG42 & $0.75\pm 0.19$ & $0.23\pm 0.06$ & $169\pm 63$ & $0.37\pm 0.12$\\
HCG68 & $0.69\pm 0.09$ & $0.50\pm 2.06$ & $200\pm 31$ & $0.22\pm 0.04$\\
NGC5171& $1.21\pm 0.05$ & $0.17\pm 0.04$ & $158\pm 10$ & $1.30\pm 0.11$\\
HCG15 & $0.85\pm 0.08$ & $0.05\pm 0.02$ & $ 81\pm 10$ & $1.48\pm 0.24$\\
NGC507& $1.34\pm 0.01$ & $0.33\pm 0.01$ & $134\pm  2$ & $2.21\pm 0.03$\\
NGC4073& $1.75\pm 0.03$ & $0.26\pm 0.02$ & $176\pm  3$ & $2.90\pm 0.05$\\
NGC4325& $0.97\pm 0.01$ & $0.39\pm 0.04$ & $122\pm  5$ & $1.09\pm 0.07$\\
NGC2563& $1.31\pm 0.05$ & $0.23\pm 0.03$ & $192\pm 11$ & $1.20\pm 0.07$\\
NGC533& $1.26\pm 0.01$ & $0.32\pm 0.02$ & $140\pm  3$ & $1.72\pm 0.06$\\
HCG51 & $1.16\pm 0.13$ & $0.59\pm 0.06$ & $186\pm 30$ & $0.92\pm 0.19$\\
HCG62 & $1.06\pm 0.02$ & $0.08\pm 0.01$ & $107\pm  3$ & $1.60\pm 0.05$\\
 \hline
\end{tabular}
\end{center}

\begin{center}
\renewcommand{\arraystretch}{0.9}\renewcommand{\tabcolsep}{0.05cm}
\caption{Properties of groups  between $0.3 r_{500}$ and $0.7 r_{500}$.}
\label{t:prop3}

\begin{tabular}{rccccccccccc}
\hline
 Name & kT  & Z & S & P \\
      & keV & $Z_\odot$&keV cm$^2$ & $10^{-12}$ dyne cm$^{-2}$ \\
\hline
3C449   &$0.67\pm 0.04$ & $0.04\pm 0.01$ & $117.\pm  10.5$ & $0.47\pm 0.05$\\ 
Pavo    &$0.56\pm 0.10$ & $0.21\pm 0.06$ & $114.\pm  35.0$ & $0.33\pm 0.09$\\ 
HCG42   &$0.75\pm 0.48$ & $0.20\pm 0.11$ & $213.\pm 156.2$ & $0.25\pm 0.20$\\ 
NGC5171 &$0.98\pm 0.12$ & $0.30\pm 0.13$ & $475.\pm 257.1$ & $0.22\pm 0.08$\\ 
NGC507  &$1.07\pm 0.01$ & $0.31\pm 0.04$ & $473.\pm  51.1$ & $0.25\pm 0.02$\\ 
NGC4073 &$1.31\pm 0.05$ & $0.10\pm 0.02$ & $247.\pm  15.8$ & $0.86\pm 0.05$\\ 
NGC4325 &$0.65\pm 0.09$ & $0.12\pm 0.03$ & $171.\pm  26.9$ & $0.26\pm 0.06$\\ 
NGC2563 &$1.08\pm 0.13$ & $0.36\pm 0.08$ & $308.\pm  71.5$ & $0.38\pm 0.08$\\ 
NGC533  &$0.92\pm 0.01$ & $0.39\pm 0.06$ & $292.\pm  32.0$ & $0.27\pm 0.03$\\ 
HCG51   &$1.24\pm 0.21$ & $0.77\pm 0.14$ & $396.\pm 130.6$ & $0.35\pm 0.16$\\ 
HCG62   &$0.78\pm 0.03$ & $0.05\pm 0.01$ & $140.\pm   9.6$ & $0.57\pm 0.04$\\ 
\hline             
\end{tabular}      
\end{center}
\end{table}

The scaled pressure profile is sensitive to the choice of the representative
temperature used to derive scalings for both $x$ and $y$ axes.  The
deviations discussed below, for example, could be compensated by typically a
20\% change in the assumption for the mean temperature.  Hydrodynamic
simulations suggest that the normalization of the pressure profile scales
well with the mass of the system (Kravtsov et al. 2006), which is therefore
used here to estimate the deviations in the weighted temperature. On the
other hand, the scaled entropy profile is quite insensitive to a choice of
the mean temperature, as a change of the scaling moves the points parallel
to the radial trend. Hence, to explain a similar relative deviation in the
entropy plot, the error in the temperature would have to be a factor of 50.

The modified entropy scaling of Ponman et al. (2003) implies that the
entropy at $0.1r_{200}$ scales as $T_w^{2/3}$. In addition, first XMM
observations revealed a similarity in the entropy {\it profiles}, scaled in
the above mentioned manner (Pratt \& Arnaud 2003, 2005). The slope of the
entropy is 1.1 outside the $0.1r_{200}$ and it is flatter inside (Pratt \&
Arnaud 2003, 2004; Sun et al. 2004). However, Pratt \& Arnaud (2005) mention
a tendency of the data to reveal a somewhat shallow slope of 1.0. The index
of 1.1 comes from 1d simulations of Tozzi \& Norman (2001). Voit (2005)
looked into the ensemble of various cosmological simulations without
feedback and derived a somewhat steeper index of 1.2. As feedback is
required in order to reproduce the observed entropy scaling (Finoguenov et
al. 2002; Voit \& Ponman 2003), the slope of the entropy profile becomes an
interesting probe relating the importance of the feedback at inner and outer
radii.

\begin{figure*}
\includegraphics[width=8cm]{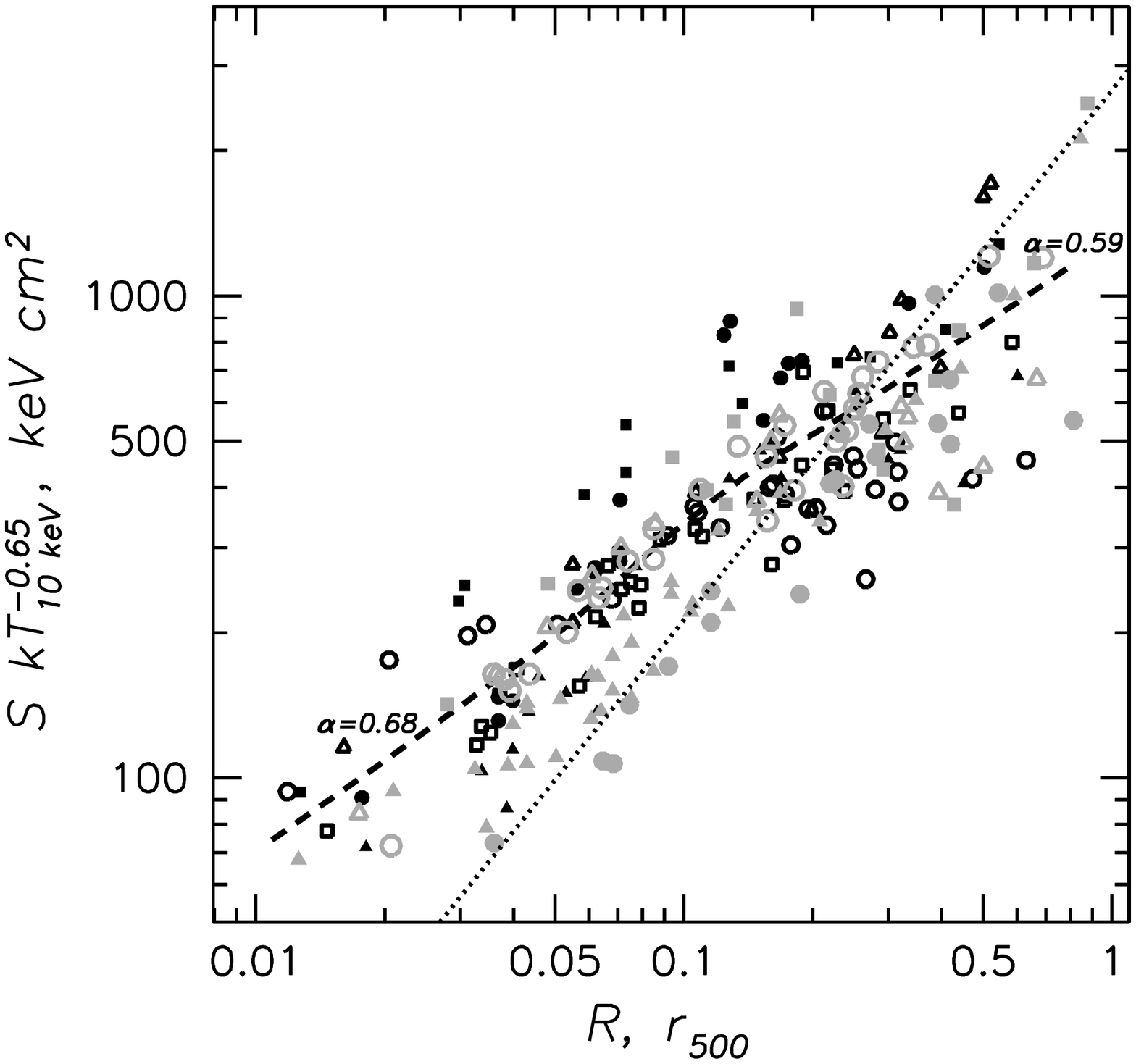}\hfill\includegraphics[width=8cm]{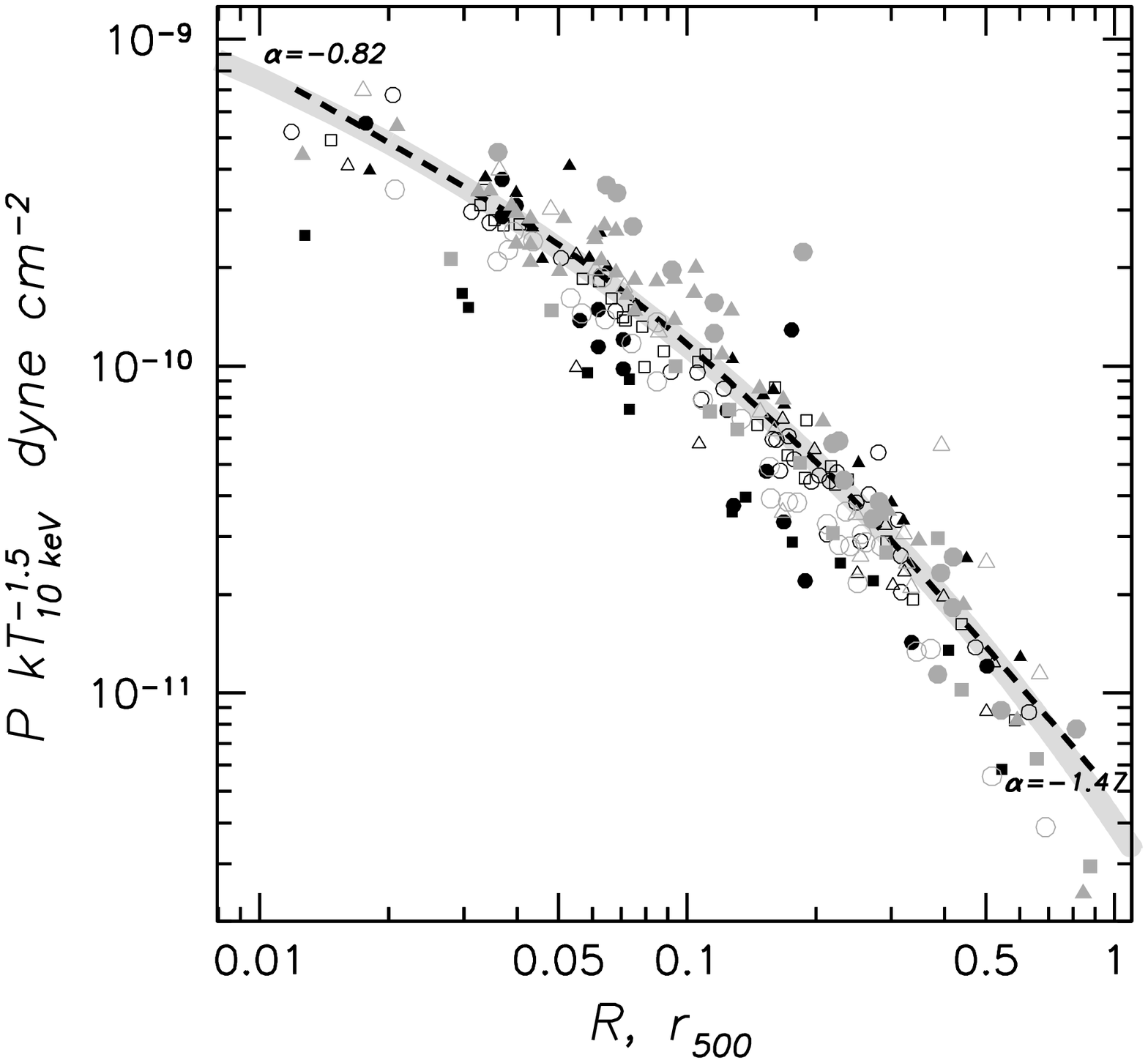}

\caption{Comparison between the entropy and pressure of the sample and
the non-parametric approximation used to study the dispersion. The entropy
and pressure points corresponding to the same group are shown using the same
symbol. The dashed line shows the results of the fit using the non-parametric
locally weighted regression method. The dotted line in the entropy panel
denotes the $S\sim r^{1.1}$ law, normalized to the results of Ponman et
al. (2003). The grey line in the pressure panel shows the S\'ersic
model with n=4, which is remarkably close to the non-parametric fit.
\label{f:comp}}

\end{figure*}

Approximating the entropy and pressure profiles using a power law and
applying orthogonal regression yields the following parameters: $S=(497\pm
5) \times ({r \over 0.2r_{500}})^{0.46\pm0.04}$ keV cm$^2$, $P=(4.5\pm0.1)
\times 10^{-11}({r \over 0.2r_{500}})^{-1.12\pm0.04}$ dyne
cm$^{-2}$. However, the power law approximation appears to be a poor fit, in
particular to the pressure profile, so we apply a more complex approach
involving non-parametric locally-weighted regression, following Sanderson et
al. (2005 and references therein). This analysis results in a non-parametric
curve, free from biases associated with model selection. The non-parametric
fit to the entropy data shown in Fig.\ref{f:comp} can be approximated with a
broken power law with inner and outer slopes of 0.78 and 0.52, respectively,
and a break around $0.5r_{500}$. As mentioned in Mahdavi et al. (2005),
there appear to be two subclasses in the entropy profiles of the groups,
where one class of profiles indeed shows the steep entropy rise in the
outskirts expected from simulations. A similar conclusion holds for this
sample and is further discussed in Osmond et al. (2006). The
characterization of the pressure profiles, shown in Fig.\ref{f:comp}, is
similar to the results of Mahdavi et al (2005) and also to that of hot
clusters (Finoguenov et al. 2005b). In the range covered by our data, a
better approximation of the pressure profile could be made with two power
laws, with slopes $-0.82$ at $r<0.1r_{500}$, $-1.47$ at
$0.2<r/r_{500}<0.7$. A steepening in the pressure, beyond $0.6r_{500}$,
present in clusters (Finoguenov et al. 2005b), also starts to be seen in our
data on groups. Remarkably, the S\'ersic (1968) law, while having only two
free parameters, can be used to reproduce the non-parametric approximation
to the pressure profile, with best-fit values for the index $n=4$ and a
normalization at $r_{500}$ of $4\times10^{-12} (T_w/10 {\rm keV})^{1.5}$
dyne cm$^{-2}$.

The amplitude of fluctuations around the best fit exceeds the effect due to
statistics, and is a measure of substructure. The average level of
fluctuations, which is 20\% for entropy and 30\% for the pressure, could be
taken as the accuracy to which the approximation of either property could be
determined at any radius. In the determination of average trends and the
scatter around them, we have excluded HCG92 and HCG68, since their
properties may not be representative of that of normal groups, for reasons
discussed below. This is a conservative approach, as their scatter around
the mean trend, reported in Tab.\ref{t:prop4} is as large as the mean
scatter for the rest of the sample.

\begin{figure*}
\includegraphics[width=8cm]{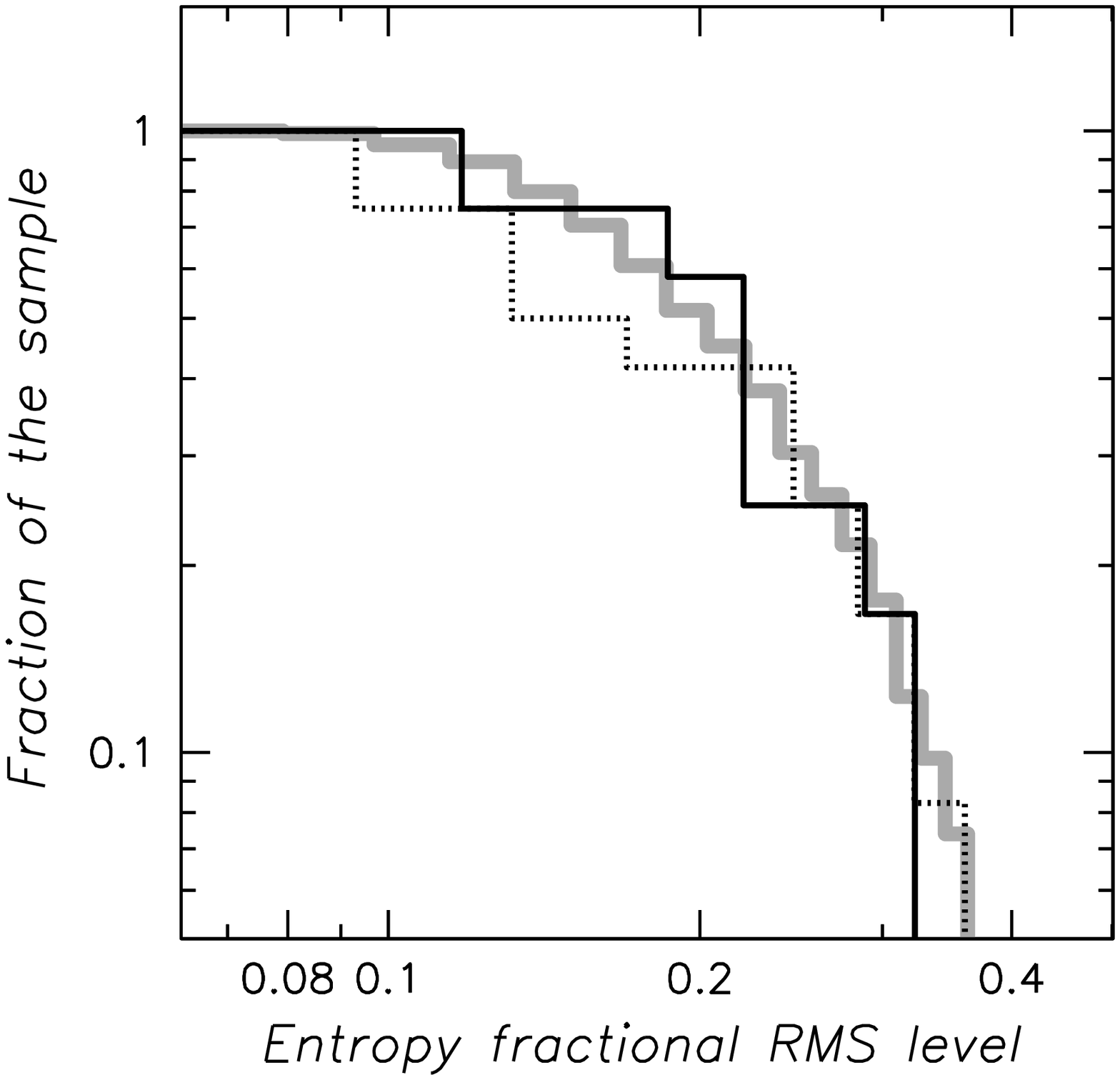}
\includegraphics[width=8cm]{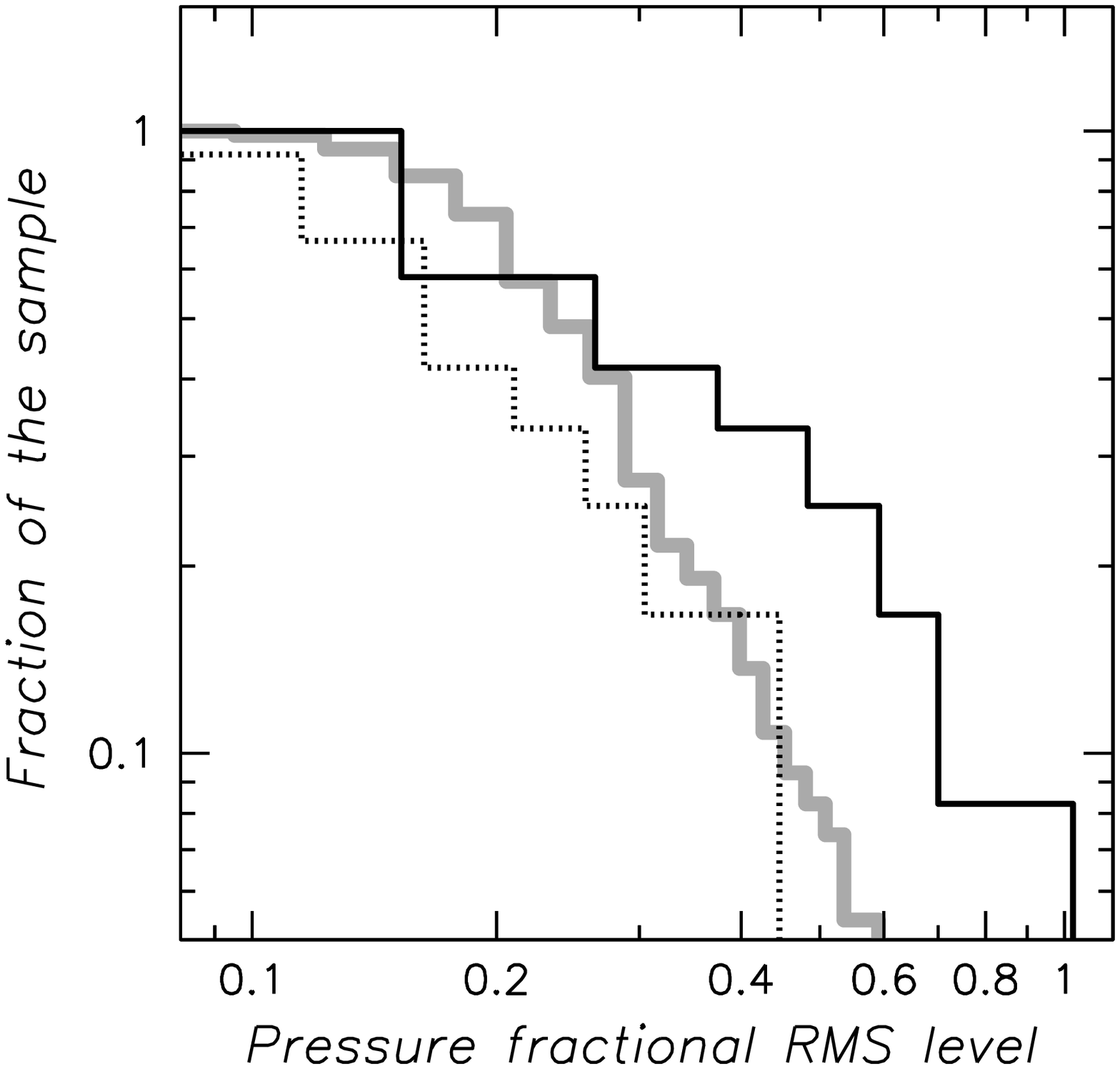} 

\caption{Cumulative distribution of groups vs fractional RMS scatter of the
  entropy (left panel) and pressure (right panel) parameter greater than the
  x-axis value. The black lines denote the results for our group sample,
  obtained using annuli and the mask sampling 2d variations in the
  temperature, marked as dotted and solid lines, respectively. The grey line
  represents the results of a 2d analysis performed on a sample of 208
  modeled clusters (Finoguenov et al. in prep.). Our group sample exhibits a
  significantly larger scatter in the pressure, while the scatter in the
  entropy is comparable to simulations.
\label{f:c2s}
}
\end{figure*}

\begin{table}
\begin{center}
\caption{Entropy and pressure fluctuations around the mean sample trend.\label{t:prop4}}
\begin{tabular}{rccccccccccc}
\hline
 Name & $\sigma S$  & $\sigma S$  & $\sigma P$ & $\sigma P$ \\
      & annuli & map & annuli  & map \\
\hline
3C449  & $0.25\pm0.03 $& $0.22\pm0.02 $& $0.03\pm0.03 $& $0.23\pm0.04$ \\
HCG 92 & $0.11\pm0.19 $& $0.20\pm0.14 $& $0.42\pm0.57 $& $0.49\pm0.21$ \\
Pavo   & $0.27\pm0.10 $& $0.19\pm0.11 $& $0.31\pm0.12 $& $0.50\pm0.27$ \\
HCG 42 & $0.05\pm0.12 $& $0.28\pm0.30 $& $0.20\pm0.47 $& $0.63\pm0.21$ \\
HCG 68 & $0.06\pm0.04 $& $0.22\pm0.19 $& $0.08\pm0.21 $& $0.32\pm0.11$ \\
NGC5171& $0.33\pm0.09 $& $0.42\pm0.14 $& $0.47\pm0.05 $& $0.22\pm0.11$ \\
HCG15  & $0.08\pm0.02 $& $0.16\pm0.04 $& $0.10\pm0.04 $& $1.08\pm0.09$ \\
NGC507 & $0.38\pm0.13 $& $0.08\pm0.01 $& $0.23\pm0.19 $& $0.08\pm0.01$ \\
NGC4073& $0.10\pm0.09 $& $0.10\pm0.02 $& $0.14\pm0.13 $& $0.12\pm0.03$ \\
NGC4325& $0.11\pm0.10 $& $0.33\pm0.16 $& $0.13\pm0.10 $& $0.42\pm0.08$ \\
NGC2563& $0.16\pm0.17 $& $0.16\pm0.06 $& $0.07\pm0.08 $& $0.15\pm0.06$ \\
NGC533 & $0.22\pm0.01 $& $0.11\pm0.01 $& $0.12\pm0.04 $& $0.11\pm0.03$ \\
HCG 51 & $0.20\pm0.09 $& $0.34\pm0.19 $& $0.28\pm0.10 $& $0.11\pm0.15$ \\
HCG 62 & $0.18\pm0.10 $& $0.20\pm0.03 $& $0.11\pm0.09 $& $0.08\pm0.02$ \\
\hline
\end{tabular}
\end{center}
\end{table}

A remarkably high dispersion of points around the mean pressure trend is
reported in Fig.\ref{f:c2s}. A similar analysis has been carried out using a
catalogue of 68 preheated clusters evolved with P3MSPH (Evrard 1988), with
the details of the analysis presented in Finoguenov et al. (in prep.). The
high dispersion in pressure could be formulated as a high fraction of the
sample with rms exceeding a 30\% value (42\% vs 22\% in the simulations) and
also having 20\% of the sample exhibiting a fractional rms exceeding 60\%,
when only 5\% are expected. The Kolmogorov-Smirnov test returns the
likelihood of the two samples being drawn from the same distribution of 1\%.
At the same time the likelihood that the rms in the entropy have a similar
origin is 97\%. A similar comparison for a representative cluster sample has
been performed in Finoguenov et al. (2005b), where no deviations between the
simulated and observed clusters were found, even though the data analysis
technique was exactly the same as that used here, and the statistics of the
observations were comparable to the current sample. Noting that the
simulations used for comparison here were simulations of {\it clusters},
rather than groups, it appears that the excess dispersion seen in the
pressure in Fig.\ref{f:c2s} is really a difference between groups and
clusters, rather than one between simulations and observations.  The good
agreement seen in the entropy, disfavours any interpretation of the large
scatter seen in groups in terms of differences in the thermodynamic history
of the gas. We suggest that the differences really lie in the dark matter
substructure, and should be further investigated in simulations. We note
that in Mahdavi et al. (2005), a similar conclusion has been derived for
some other groups individually, so we think that our result is
representative.

This result is somewhat surprising, given that the CDM scenario predicts
little dependence of the subhalo mass function on the mass of the host. A
comparison with the data on the lowest mass scale, probed by observations of
the nearby galaxies, reveals an opposite problem, as too few dwarf galaxies
are found (e.g. Klypin et al. 1999). While a number of baryonic effects
could be used to cure the problem on dwarf regime, these observations may
shed light on the nature of the dark matter. Here we discuss a possible
explanation within the framework of the conventional $\Lambda CDM$ model. In
general, at an enhanced level of the entropy of the gas, typical of galaxy
groups, one expects to diminish the effects of substructure (e.g. Kay et al.
2004), unless it retains its own gas. We stress once again, that difference
in the average trend, also examined in Fig.\ref{f:c2s} using the annuli, can
not account for the observed two-dimensional scatter. In our case, the most
deviant points on the maps could often be associated with the location of
major galaxies, yet the effect is not sufficient to disturb the entropy
distribution. As at the moment we do not have simulations of groups to
compare with, we postpone resolving this issue to the future work.

Another potentially important effect arises from the narrow redshift range
of the groups in our sample. There is a known difference in the abundance
of groups between the northern and southern hemispheres, associated with
the presence of a large scale structure at $z\sim0.02$ (Boehringer et
al. 2002).  Thus, further X-ray observations of a more distant group
sample, such as available from the ROSAT 400 square degree survey (Burenin
et al. 2006), are needed to study the influence of LSS on these
conclusions.

\subsection{Velocity structure of groups}\label{s:opt}

\begin{figure*}

\includegraphics[width=5.5cm]{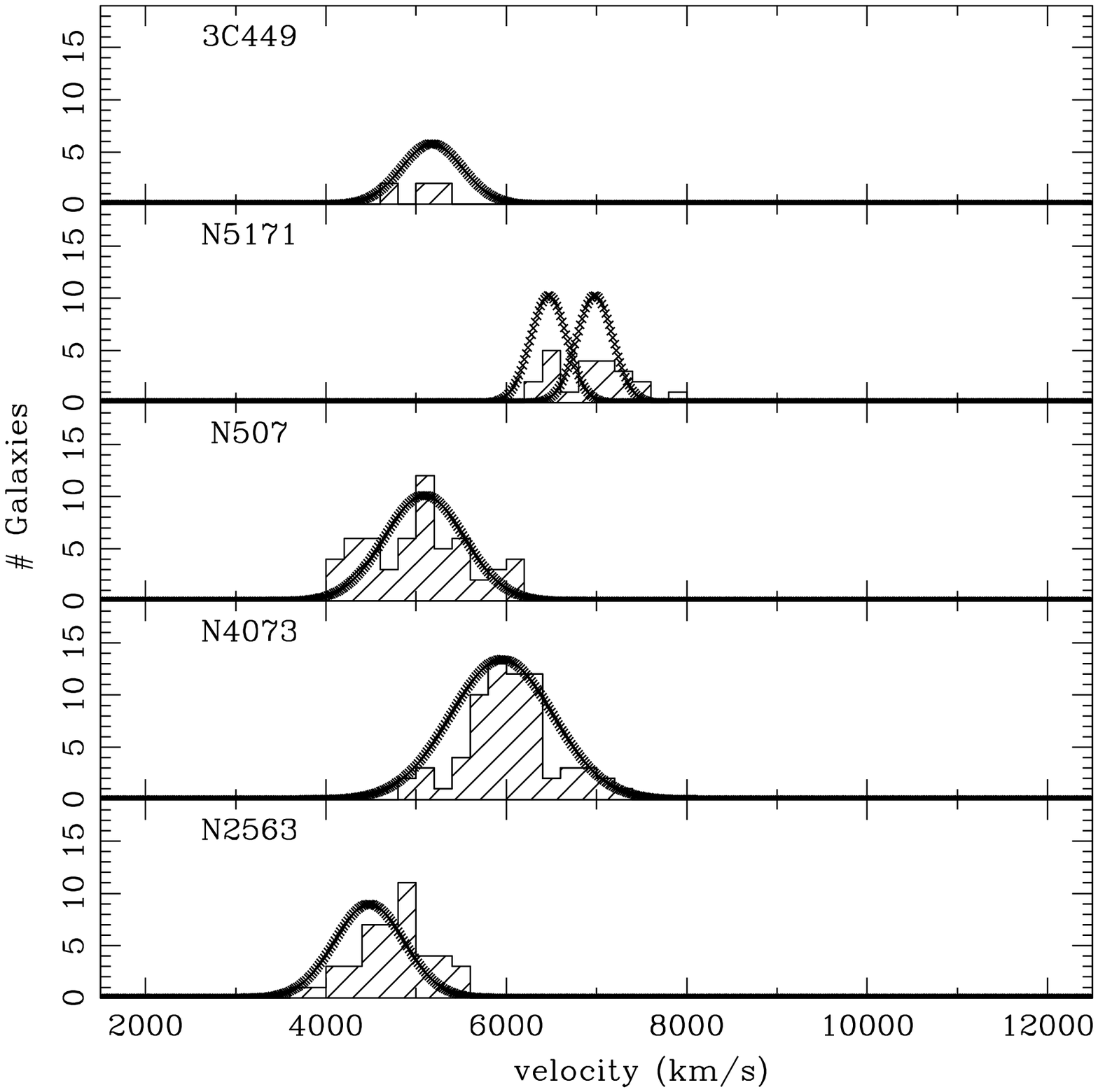}
\includegraphics[width=5.5cm]{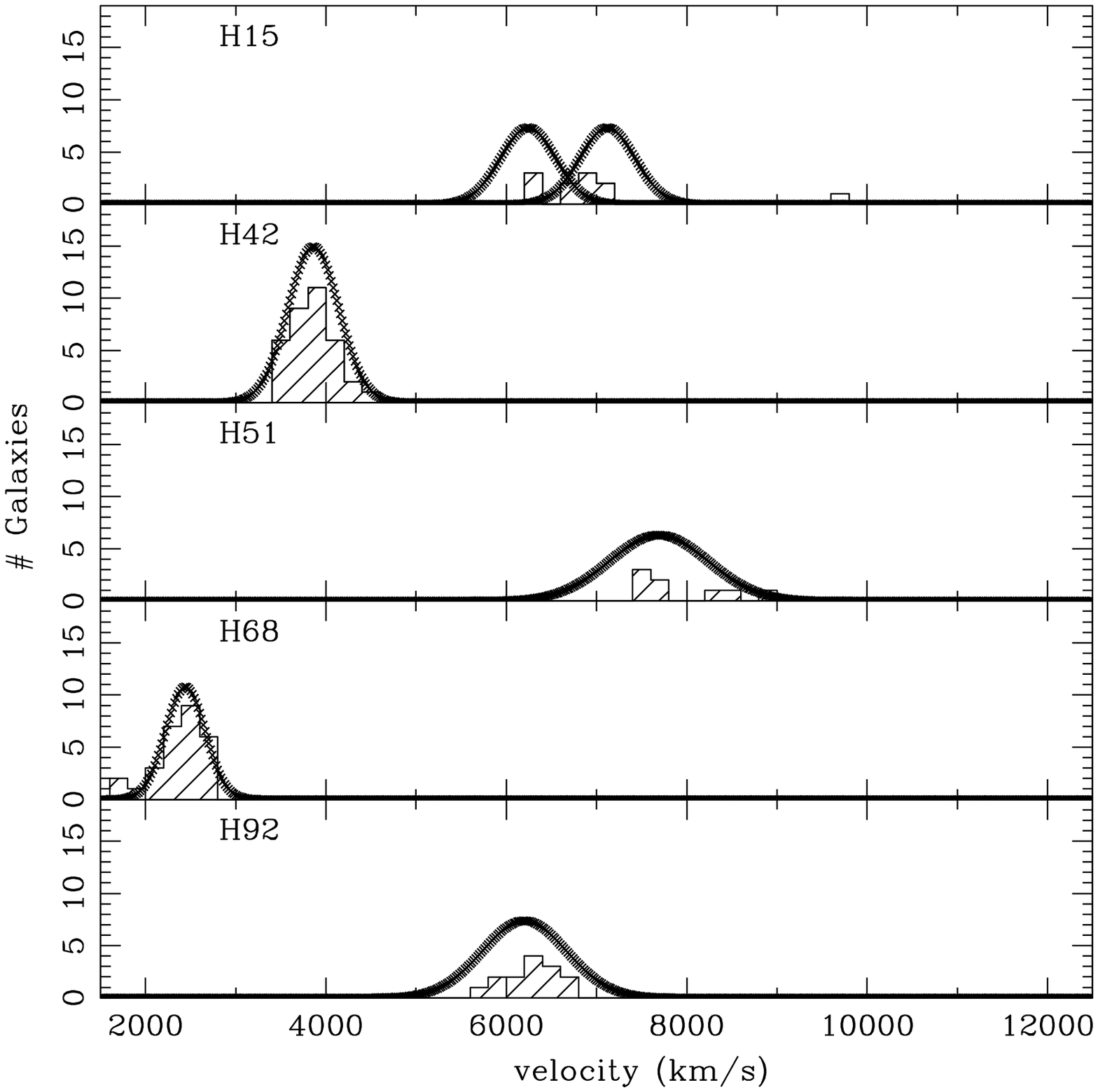}
\includegraphics[width=5.5cm]{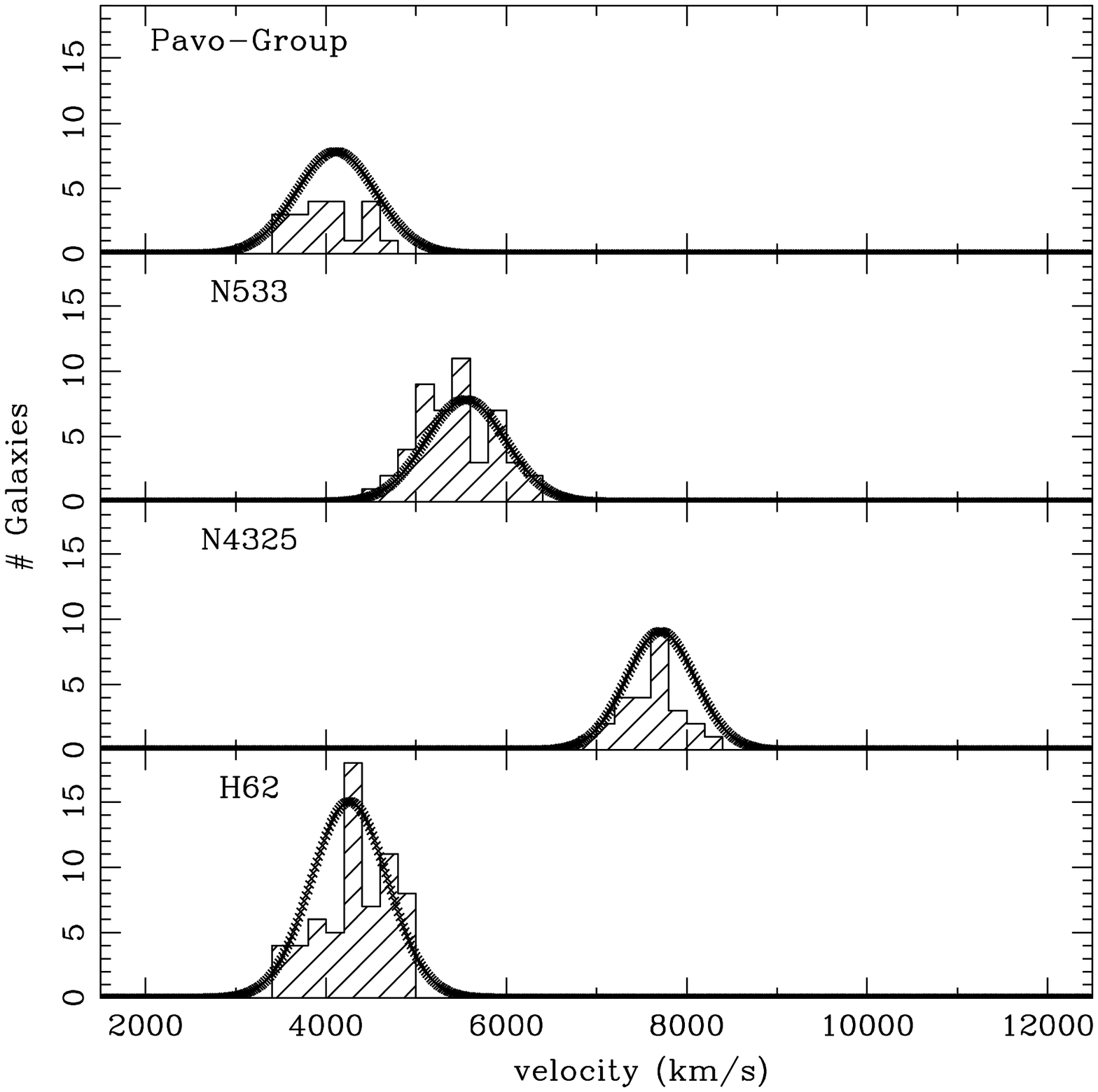}

\caption{Galaxy velocity histograms for the sample. The panels display the
  results obtained using the binning of the data of 200 km s$^{-1}$.
\label{f:opt}}

\end{figure*}

If, as suggested above, our group sample exhibits the presence of
substructure in the dark matter, evidence of this might also be apparent
in the velocity histograms of galaxies. To explore this, we collect 
them together in this section, although most of these histograms have
been already published elsewhere (Zabludoff \& Mulchaey 1998).

Fig.\ref{f:opt} displays the velocity histograms of the sample, using a bin
size of 200 km/s and including galaxies within the 1.2 Mpc of the group
centre. A typical error on estimating the velocity is around 80--150
km/s. Evidence for substructure is present for 3C449, NGC5171, NGC507 and
HCG15, although its exact quantification is difficult in most cases, due to
the poor statistics. NGC4073, NGC2563, NGC533 and HCG62 appear to be regular
groups. Some hints of substructure seen in the HCG62 histogram have been
robustly identified through a refined method by Zabludoff and Mulchaey
(1998). The available quality of the data is not sufficient for
characterizing the degree of substructure to better than 20\%, and a
dedicated follow-up program is underway, the results of which will be
reported elsewhere (Zimer et al. in prep.). With the currently available
data, the conclusion is that, while in some cases (discussed in section 4
below) a detection of substructure provides supporting evidence of a link
between the observed properties of the gas and the underlying dark matter
distribution, it is not possible to derive quantitative conclusions without
a much larger spectroscopic database on group members. Typically, a minimum
of 100 group members with redshift measurements is required for such a
study.

The brightest group member is always an elliptical, except for HCG92, where
it is a lenticular. However, a complex galaxy (e.g. a double galaxy) in the
centre is present for HCG 68, HCG 92, NGC 4073, Pavo, 3C449, HCG 51, NGC
507, HCG 62. 
Pavo, HCG51, NGC4073 contain more than one spiral galaxy.

\section{Individual properties of the groups}\label{s:ind}

We will follow a similar scheme when presenting the results for every
system. As described in captions to Fig.\ref{r:3c449}, from top to bottom,
left panels show the profiles of entropy, pressure, temperature and Fe
abundance\footnote{The two dimentional information, related to the analysis
reported in this paper, is released under http://www.mpe.mpg.de/2dXGS/
homepage}. The results obtained using annuli are shown in grey. The results
from two-dimensional analysis (black crosses) are both shown as maps in the
central panels and are converted into the profiles by plotting the observed
values vs the distance of the extraction area from the centre. The dashed
line in the entropy panel shows $S\propto r^{1.1}$, expected from
simulations, normalized to the universal entropy scaling relation of Ponman
et al. (2003). The non-parametric model for entropy and pressure are plotted
as dashed line in the corresponding panels.  Entropy and pressure maps are
displayed as their ratios to the non-parametric model.  From top to bottom,
right panels show the results of image analysis: entropy, pressure,
temperature maps, and surface brightness in the 0.5-2 keV band, where the
temperature has been estimated through the hardness ratio of the 0.5--1 and
1--2 keV bands. Coordinates on the images are in units of $r_{500}$. The
iron abundance is given in the photospheric solar units of Anders \&
Grevesse (1989).

Within the scenario of smoothness of accretion (e.g. Voit et al. 2003;
Borgani et al. 2005), lumpy accretion should lead to lower entropy in group
outskirts. At the same time, this would produce a large degree of variance
in the central regions. While this is true for 3C449, a subsample of systems
with lower entropy at outskirts, which includes 3C449, Pavo, NGC5171,
NGC4073, NGC533, NGC4325 and HCG62, does not encompass all deviant systems.
Some systems with high entropy at outskirts, such as HCG51, HCG42, have
large degree of substructure. In addition, the flat temperature profile of
HCG51 and HCG42 is a rather striking feature, possibly suggesting a major
merger. Thus, while mergers seem to account for all systems of high
dispersion, they either result in flat temperature profiles or in flat
entropy profiles, probably depending on the mass ratio of the merger.  The
low entropy at outskirts of some of the groups may indicate an infall of the
gas which corresponds to the early stage of a merger. We associate flat
temperature profiles with the late stages of mergers.


\subsection{3C449}

{\it As the number of figures is 176 in this paper, they are omited in
astro-ph version, please follow the pointer in the astro-ph abstract}. The
surface brightness of 3C449 (UGC 12064) is elongated in the NE-SW direction
on scales of $0.3r_{500}$ and in addition the NE sector has an enhancement
in the form of a long tail or an arc, extending from the centre to the NE
and turning north at $0.1r_{500}$.  The spectral analysis reveals that the
SW sector exhibits both higher pressure and lower entropy relative to the
corresponding model describing the profiles of 3C449. The NE sector has
lower entropy and higher metalicity. In the entropy plots, the system
appears as quite deviant, exhibiting practically constant entropy level from
$\sim0.1r_{500}$ to $0.7r_{500}$.  The temperature only drops at the very
centre of the system. The central drop in the Fe abundance could be an
artefact due to large temperature diversity in the centre (Buote 2000).
\subsection{HCG15}

The wavelet analysis of HCG15 reveals an abundance of structure on scales
from point sources to $0.3 r_{500}$. All of
the substructure associated with the extended source is centered on the
member galaxies of HCG15. On the larger scale, only the surface brightness
map was obtained. The spectral analysis using annuli reveals a typical
pressure profile. The entropy profile is approaching the scaling predictions
at $0.2r_{500}$. On the scaled entropy and pressure profiles HCG15 reveals
the component at $0.1r_{500}$ distance to the centre, associated with one of
the galaxies. An asymmetry in the X-ray appearance of HCG15 is entirely due
to the member galaxies, which also is responsible for the non-monotonic
radial behavior of the entropy in the analysis using annuli. The iron
abundance is typically below 0.1 solar, which is surprisingly low at
locations associated with the galaxies. The available statistics of the data
preclude a more detailed spectral investigation. The outmost rings do not
have sufficient flux from the source to constrain the temperature and are
not shown in the plots. The observed part of the group is in good agreement
with the scaling, making the case of HCG15 different from that of HCG 68 and
HCG 92.

\subsection{NGC2563}

The entropy map appears to be quite regular. Only seven regions are
available for detailed spectroscopy. The gas properties are traced to
$0.6 r_{500}$. The temperature is 0.9 keV in the centre, rising to 1.6 keV by
$0.1r_{500}$ and then dropping. The iron abundance is 0.6 times solar at the
centre, dropping to 0.2 solar value at outskirts. The entropy is high at the
centre, approaching the sample average trend at $0.3 r_{500}$. Thus, the low
X-ray luminosity of NGC2563 is associated with an inflated group core,
while the bulk properties of its IGM  are typical, which is also
seen in the pressure plot. On the other hand, the feedback effects inside
$0.3r_{500}$ are remarkable. The entropy shows a shelf between the 0.1 and
0.3 $r_{500}$ and at the same time, the inward raise in the pressure from
$0.3r_{500}$ becomes immediately flatter, which for a similar mass profile
is expected when gas becomes lighter. Such behavior of pressure supports the
idea that the high entropy level is a dominating feature of the gas between
0.1 and 0.3 $r_{500}$.  At very centre, however, the entropy of NGC2563
becomes typical.

\subsection{HCG42}

For the central elliptical the temperature is 0.75 keV and the iron
abundance is 0.5 solar. The temperature slowly rises outwards reaching 0.85
keV. On the pressure map, the eastern part appears enhanced, indicating
substructure. The eastward part of HCG 42 has lower entropy compared to the
model, which is also associated with lower metalicity. On the entropy plot,
HCG42 appears to approach the general behaviour for groups starting at
$0.2r_{500}$, meaning that HCG42 should be considered as a virialized group,
though affected by additional feedback effects. Apart from the substructure
to the east, on the pressure plot HCG42 also appears as a normal group. We
see no substantial deviations from Gaussianity in the the velocity histogram
of HCG42.

\subsection{NGC507}

Within the central $0.2r_{500}$ NGC 507 exhibits an elongation in the
surface brightness in the NW-SE direction, which is particularly outstanding
in the NW, where a spectral analysis reveals lower entropy, higher pressure
and higher metalicity. Two central galaxies are seen in NGC507, and the
velocity histogram supports the idea of substructure. The temperature
structure needs more than one component for the region centered on the
galaxy. The temperature lies mostly in the 1.0--1.3 keV range, with iron
abundance exceeding 0.5 solar, dropping to 0.35 in the outermost region. On
large scales, the system is relaxed. The gas properties are traced out to
$r_{500}$ and are consistent with standard entropy scaling. 
In the outskirts the pressure
profile is very steep compared to average trends for the groups. A rise
of both entropy and pressure to the scaling at the outmost point may not be
representative for this system, since only a small part of the system is
observed at this radius, as a result of the instrumental setup of the
observation, in which the centre of NGC~507 was shifted to the
corner. Higher entropy and lower pressure are usually found in simulations
in the direction perpendicular to the alignment of filaments (Kravtsov,
A. 2004, private communication).

\subsection{HCG 68}

HCG 68 is one of the most underluminous systems in the sample. HCG 68
reveals higher entropy to the east and north of the centre, as seen in both
the hardness ratio based analysis and direct spectral fitting. The
temperature behavior is nearly isothermal at the $\sim 0.64$ keV
level. Hotter temperature zones are seen to the north, typically on the 0.75
keV level with a maximum of 0.9 keV. The entropy of HCG68 is traced to
$0.3r_{500}$ where it is still higher than predictions from scaling. The
deviations from the scaling are the largest for this system and are similar
to HCG92: very high entropy and very low pressure of the gas. As appears to
be the case for HCG92, major galaxy merger events within a low-mass group is
a probable explanation for the presence of the X-ray emission.  From the
three galaxies located at the group's center,
at least two are spirals. On the other hand, the low
metalicity of the gas restricts  the contribution from 
stellar mass loss. If the origin of the hot gas is similar to HCG~92 --
heating of the HI -- the low metalicity should match the metalicity of the HI
phase. The velocity histogram exhibits skewness, indicative of infall,
while the mean temperature appears to be too hot compared to expectations
for both the measured velocity dispersion and the normalization of the
pressure profile.

\subsection{Pavo}

The central part of the group contains two major galaxies. The coordinate
grid is centered on the elliptical and the spiral is located at (0.35,0.30)
and is associated with an enhancement in the X-ray surface brightness. In
X-rays, there is a bridge between the two major galaxies, with a possible
sign of interaction on the side of the spiral. The temperature around the
main elliptical is 0.9 keV and declines outward reaching the $\sim 0.5$ keV
level. The spiral galaxy exhibits a low temperature of emission. Pavo groups
exhibits a quite low level of entropy between $0.3-0.7r_{500}$ compared to
the average trend. The pressure profile shows an enhancement at these radii,
associated with the location of the spiral.

\subsection{HCG 92}

A detailed analysis of the XMM-Newton observations of the HCG~92 is
presented in Trinchieri et al. (2005). A remarkable feature of HCG~92
consists in the shock heating of HI (Trinchieri et al. 2003). The X-ray
surface brightness enhancement, which is extended in origin, as revealed by
Chandra, is seen in Fig.\ref{r:h92} as an extension from the centre to 0.1
towards the south in the surface brightness. The most surprising finding of
our comparative analysis is that, despite a completely different origin for the
X-ray gas, its entropy and pressure does not deviate from the mean trend for
groups. 

The average temperature of the IGM in HCG~92 is $\sim0.6$ keV, with hotter
regions located in the south-east. The element abundance profile rises with
radius, indicating a lower metalicity for the recently heated gas.  However,
a downward bias in derived abundance due to complexity of the temperature
structure (Buote 2000), is also possible. The entropy and pressure profiles
are traced out to $0.3r_{500}$, where they start to deviate from the mean
trend seen for groups. Such deviations could potentially be used to
delineate the low-mass groups with high X-ray luminosity, caused by merger
events.

At a distance of $0.3r_{500}$, there are zones of
enhanced pressure and temperature, which have low metalicity. We associate
these zones with the heating of the infalling low-metalicity material.

\subsection{MKW4}

The level of fluctuations in the entropy and pressure in IGM of MKW4 is
rather moderate, indicating that the system is close to hydrostatic
equilibrium. In the centre, the temperature is 1.6 keV and the iron
abundance of 1.5 solar. The temperature initially rises to 2.2 keV, then
declines. Iron abundance drops to the 0.2 solar level. Outside $0.1r_{500}$
the entropy of MKW4 is lower than predicted by the scaling. The largest
deviations from spherical symmetry are seen in entropy $0.2r_{500}$ west
from the centre and could be explained as the fossil record of a previous
minor merger. The temperature profile presented here agrees with the Chandra
results in Vikhlinin et al. (2005), the Chandra/XMM analysis of Fukazawa
et al. (2004) and the ROSAT/ASCA results in Finoguenov et al. (2000).
Deviant results are presented in O'Sullivan et al. (2003), based on a
similar XMM dataset, but erroneously taking the MKW4 emission within the XMM
FoV for the 'soft excess' associated with foreground.

\subsection{NGC5171}

A detailed analysis of NGC5171 is presented in Osmond et al. (2004), where
the appearance of the group has been explained by the merger of two groups,
with a distant group appearing in projection in the south-east, which in our maps
appears as the most deviating point in both the entropy and pressure maps.
Entropy to the north is low, and represents the stripping tails of the
interaction. Also, an elongation in the pressure to the north suggests that
the dark matter potential of the infalling group has not been destroyed
yet. The global properties of the system on large scales are quite normal,
in both entropy and pressure, indicating that although the interaction has
an effect on the luminosity of the object, it does not affect the bulk of
the gas outside $0.3r_{500}$. The overall pressure level is low, which
indicates that the temperature of the system used for scaling has been
boosted by 20\%. This corresponds to differences between the adopted
temperature measured between 0.1 and $0.3 r_{500}$, where in fact the
interaction between the groups occurs, and the temperature measured between
0.3 and $0.7r_{500}$. The element abundance is low in most regions and is
poorly constrained.

\subsection{HCG62}

Within $0.1r_{500}$, HCG62 exhibits an interesting bubble-like temperature
structure and cool extensions to the north, which have also been reported in
the Chandra observations (Vrtilek et al. 2002). On larger scales, the
distribution of IGM properties is very symmetrical. Temperature is 0.8 keV
at the centre rising to 1.3 keV before starting to fall back to 0.8 keV. On
the entropy plot HCG62 exhibits a plateau between 0.1 and $0.3r_{500}$ with
entropy lower than the scaling relation predicts, but then starts to rise
again. We associate such behavior with the infall zone, also evident in
the velocity histogram. HCG 62 provides an important example; the
entropy profile is flat within a substantial part of the group and is at the
level of 100 keV cm$^2$, yielding small values of beta in the surface
brightness analysis, and yet it can hardly be the result of feedback, as
the entropy of the gas is lower than seen on average. The overall level of
the pressure is high in HCG62, probably due to an underestimate of the scaling
temperature by 20\% due to low entropy inclusions.

\subsection{NGC4325}

NGC4325 deviates strongly from the mean trend of the sample, e.g. the
surface brightness profile for this group is very peaked, at odd with the
generally flatter profiles of other systems. As a result, we obtain large
dispersion values for it. Within the central $0.2r_{500}$ the entropy
profile follows closely the expectation for cool core systems, and there is
a corresponding pressure enhancement.  These measurements reflect a very
peaked surface brightness profile for this group, compared to other systems.
Outside the central $0.2r_{500}$, the system regains a typical entropy and
pressure level. The Fe abundance steadily declines with radius. The high
metalicity ring at $0.4r_{500}$ distance from the centre is not very
significant. The temperature behaviour is typical, consisting of initial
raise from 0.8 keV, flattening at 1 keV and a subsequent decline reaching
0.6 keV at $r_{500}$.

\subsection{NGC533}

The IGM of NGC~533 shows typical profiles for both pressure and entropy.  The
gas is traced out to $0.7r_{500}$. The temperature raises from 0.9 at the
centre to 1.4 and then falls to 0.8 keV at the limit of
observation. The velocity histogram shows two peaks, the location of the two
subgroups corresponding to the elongation in the surface brightness in the
NE--SE direction.  In the maps, a distinct difference between north-eastern
and south-western parts of the group is seen. The north-east is hotter,
mostly due to the high entropy, with surprisingly higher Fe abundance. We
associate these features with incomplete mixing after the merger of the
two groups, which is also responsible for the anisotropy in the velocity
histogram of galaxies. The pressure plot indicates that the scaling
temperature of the system has been boosted by 20\%. This indicates that it
is the NE part of the group that is affected by merging.

\subsection{HCG51}

The statistics of the XMM observations of HCG~51 are rather poor, as only 5
ksec was left after the flare cleaning. However, a number of interesting
features are seen. On average, the temperature exhibits an isothermal
behavior to $0.7r_{500}$, and a lack of metalicity gradient, confirming the
results of the study of Finoguenov \& Ponman (1999) and is suggesting a
recent major merger.  Hotter zones are seen to the south-west at
$0.2r_{500}$ and to the west at $0.5r_{500}$. The iron abundance is also
enhanced there. On small scales, the surface brightness reveals a number
of asymmetries, which we were not able to resolve spectroscopically. The
observed picture suggests incomplete mixing after a recent merger. HCG
51 is also a very dense system in the optical. On the entropy profile, the
system reveals quite a high entropy level in its outskirts. The overall pressure
level is low, suggesting that the scaling temperature has been boosted by
10\%.

\section{Summary}\label{s:sum}

We have performed a detailed study of the IGM in a representative sample of
groups of galaxies observed by XMM-Newton, which has been primarily drawn
from the sample of Mulchaey et al. (2003). Comparing the entropy and
pressure profiles of these groups with typical cluster profiles, scaled
using the prescription of Ponman et al. (2003), we perform a non-parametric
orthogonal regression analysis and obtain the typical entropy and pressure
profile of the sample. In comparison to the commonly adopted $r^{1.1}$
behaviour for the entropy, the averaged profile for the groups is flatter,
$r^{0.6-0.7}$, which is due to two effects: feedback in the group centres
and the influence of infall in the outskirts. The averaged pressure profile
has a slope of $-0.8$ within $0.1r_{500}$, and gradually steepens, reaching
a slope of $-1.4$, which at the same radii is still shallower compared to
the value of $-2.5$ reported in the similar analysis of clusters of galaxies
(Finoguenov et al. 2005), and the overall mean pressure profile is well
described by a S\'ersic law with $n=4$.

The analysis using annuli may result in a somewhat steeper entropy profiles,
compared to the results reported here. As discussed in Xu et al. (2004), a
two-temperature model accounting for mixing of components with different
entropy, assuming they are in the pressure equilibrium, also results in a
flatter entropy profile. So, it might be that a two-dimensional modelling
starts to resolve those components and reveals on average higher entropy at
the center, just like the two-temperature approach. There are however,
several pitfalls here. One is that we cannot verify this by performing a
two-temperature fit, as the binning of the regions would suffer from high
requirements on the statistics of the spectra imposed by a two temperature
model. On the other hand an assumption of pressure equilibrium is not often
justified, as shock wave driven by AGN are often observed as in e.g. M87
(Forman et al. 2005) and the Perseus cluster (Fabian et al. 2005).

An abundance of substructure has been identified in the sample, most of
which we associate with the debris of recent mergers, usually seen as
incomplete gas mixing. We also identify cases where the gas simply
traces the dark matter substructure. In most cases we see also anisotropy in
the velocity distribution of the galaxies, though our statistics
are generally poor.

Our two-dimensional study yields high dispersion values for
the pressure, compared to numerical simulations, which describe well
the trends observed in clusters of galaxies. On the other hand, the scatter
in the entropy is found to be as predicted in the simulations and is similar
to that in clusters. This could either mean that the subhalo mass function
in groups is different to that of the clusters, or that the substructure is
better seen over the fainter mean trends typical of groups. We also mention
that the sample selection correspond to a single large scale structure and
studies covering larger redshift depth are required to firmly establish our
findings. 

We find that the state of the IGM resulting from galaxy merging, as
observed in HCG~92, resembles closely the scaled properties of luminous
galaxy groups. Thus, even using entropy and pressure it will be hard to
identify the low-mass groups upscattered in their $M-L_x$ relation due to
merger events, as discussed in Stanek et al. (2006).

\section{Acknowledgments}
The XMM-Newton project is an ESA Science Mission with instruments and
contributions directly funded by ESA Member States and the USA (NASA). The
XMM-Newton project is supported by the Bundesministerium fuer Wirtschaft und
Technologie/Deutsches Zentrum fuer Luft- und Raumfahrt (BMWI/DLR, FKZ 50 OX
0001), the Max-Planck Society and the Heidenhain-Stiftung, and also by
PPARC, CEA, CNES, and ASI. AF acknowledges support from BMBF/DLR under grant
50 OR 0207 and MPG. AF\&MZ thank the Birmingham University for hospitality
during their visits. The authors thank the referee for a careful reading of
the manuscript and providing useful suggestions on presentation of the
material.

\label{lastpage}

\end{document}